\documentclass[twocolumn,showpacs]{revtex4}
\usepackage[english]{babel}
\usepackage[ansinew]{inputenc}
\usepackage[T1]{fontenc}

\usepackage{graphicx}
\usepackage{tikz}
\usetikzlibrary{patterns}
\usepackage{amsmath,bm}
\usepackage{amssymb}
\usepackage{mathrsfs}

\usepackage{braket}
\usepackage{comment}

\newcommand{\abs}[1]{\left\lvert #1 \right\rvert}
\newcommand{\tuborg}[1]{\left\{ #1 \right\}}
\newcommand{\af}[1]{\left( #1 \right)}
\newcommand{\kantpar}[1]{\left[ #1 \right]}

\newcommand{\intpar}[1]{\! \mathrm{d}{#1} \ }
\newcommand{\intparNS}[1]{\! \mathrm{d}{#1}}

\newcommand{\defi}{\equiv}

\newcommand{\Op}[1]{\hat{#1}}

\newcommand{\CC}{* }

\newcommand{\pardd}[3][]{\frac{\partial^{#1} #2}{\partial #3^{#1}}}

\newcommand{\Latt}{\mathrm{lat}}
\newcommand{\Identity}{\Op{\mathbf{1}}}

\newcommand{\TB}{\mathrm{TB}}
\newcommand{\Int}{\mathrm{int}}

\newcommand{\Eff}{\mathrm{eff}}

\newcommand{\Max}{\mathrm{max}}

\newcommand{\imag}{\mathrm{i}}
\newcommand{\Exp}[1]{\mathrm{e}^{#1}}

\newcommand{\kommu}[2]{[ #1 , #2 ]}

\DeclareMathAlphabet{\mathboring}{OT1}{cmss}{bx}{it}

\newcommand{\Oddity}{p}
\newcommand{\Integers}{\mathbb{Z}}

\newcommand{\ExcNum}{n}
\newcommand{\SubExcNum}{m}
\newcommand{\WaveNum}{k}

\begin{document}

\title{Born-Oppenheimer description of two atoms in a combined oscillator and lattice trap}

\author{Ole S\o e S\o rensen}
\affiliation{Lundbeck Foundation Theoretical Center for Quantum System Research, Department of Physics and Astronomy, University of Aarhus, DK-8000 {\AA}rhus C, Denmark}
\author{Klaus M\o lmer}
\affiliation{Lundbeck Foundation Theoretical Center for Quantum System Research, Department of Physics and Astronomy, University of Aarhus, DK-8000 {\AA}rhus C, Denmark}

\begin{abstract}
We analyze the quantum states of two identical bosons in a combined harmonic oscillator  and periodic lattice trap in one spatial dimension. In the case of tight-binding and only nearest neighbor tunneling, the equations of motion are conveniently represented in the momentum representation. We show that in the case of strong attraction between the particles, the different time scales of relative and center-of-mass motion validate a separation of the problem similar to the Born-Oppenheimer approximation applied in the description of electronic and nuclear motion in molecules. 

\end{abstract}
\maketitle

\section{Introduction}

Cold atoms in periodic potentials formed by standing wave laser beams offer a test bench for a multitude of physics phenomena ranging from single particle band structure and Bloch oscillations over artificial gauge potentials  to many-body  transport properties and phase transition dynamics \cite{bloch2005ultracold,RevModPhys.80.885,PhysRevLett.102.230601}. The system offers control over particle density and  tunneling and interaction strengths, and read-out is accommodated by fluorescence detection of the atoms, either in the far field interference after release from the lattice potential  \cite{greiner2002quantum} or within the lattice \cite{bakr09,sherson-2010-467}.

In a periodic potential, the quantum state of two atoms is separable in total and relative coordinates, and one has analytical access to states bound by attraction between the atoms and also to states held together by the combination of a repulsive interaction and the band structure due to the lattice potential \cite{winkler-2006-441,PhysRevA.76.023607,0953-4075-41-16-161002,PhysRevA.81.042102}.
A number of publications have dealt with the separation of the center-of-mass and the relative motion in degenerate quantum gases
\cite{PhysRevA.81.042102,PhysRevA.81.043609,PhysRevA.78.033611,PhysRevA.82.043615,PhysRevA.72.043620}.
Recently, we have \cite{PhysRevA.85.043617} investigated the lattice system with periodic boundary conditions in the tight binding approximation and found that an accurate diagonalization of the many-body Bose-Hubbard Hamiltonian leads to eigenstates which can be recognized as superpositions of translated replicas of a single bound composite many-body state. The phase factors chosen for this superposition govern the center-of-mass momentum of the atomic ensemble, while the relative motion of the atoms is accounted for by the bound composite quantum state. In \cite{PhysRevA.85.043617} we verified that for sufficiently strong attraction, the motion within the composite object occurs on a more rapid time scale than the center-of-mass motion, justifying the separation of the two degrees of freedom for both the ground state and the lowest excited states of the system.

It is the purpose of this manuscript to investigate the validity of a separation of the relative and center-of-mass motion for the problem of two attractively interacting bosons in an optical lattice similar to the Born-Oppenheimer approximation used in molecular chemistry. In addition to the lattice we apply a confining harmonic potential, so that the separation of coordinates is not guaranteed by symmetries and conservation laws but has to be justified by a physical argument valid only in appropriate limits.

 In Sec. II, we present the Hamiltonian describing our system, and we derive an expression for the Hamiltonian in continuous quasi-momentum space  rather than in the discrete lattice position space. In Sec. III, we introduce our separation of the problem in center-of-mass and relative quasi-momentum coordinates, and we identify the symmetries and boundary conditions of the states on the suitable reciprocal lattice. In Sec. IV, we motivate the Born-Oppenheimer separation of the problem, which leaves us with two one dimensional eigenvalue equations. In Sec V we present numerical solutions to the problem, that we compare with solution of the full two-body Schr\"odinger equation. Structures in the solutions and spectra can be interpreted via the Born-Oppenheimer separation, which also offers analytical approximations in the different parameter limits.

\section{Lattice Hamiltonian}

\subsection{One-body Hamiltonian and Wannier states}

We consider a  particle moving in a sinusoidal potential, so that the Hamiltonian can be written
\begin{align}
  \Op H_\Latt &= \Op P^2 + V_0 \sin^2(\pi \Op X)
\label{eq:EnergyEigenequation}
\end{align}
where the scaled position and momentum operators have the dimensionless commutator
\begin{align*}
  \kommu{\Op X}{\Op P} &= \imag.
\end{align*}
Since the potential is periodic with unit period, Bloch's theorem ensures that we can choose energy eigenstates
  \begin{align*}
    \Op H_\Latt \ket{\psi_q^{(n)}} &= E_q^{(n)} \ket{\psi_q^{(n)}}    
  \end{align*}
with quasi-momenta $q \in ]-\pi , \pi]$ and band indices $n = 0,1,\ldots$. Another basis---the Wannier states---can be obtained as the Fourier transform over a single Brillouin zone of the eigenstates
\begin{align*}
  \ket{w^{(n)}_{k}}
  =  \frac{1}{\sqrt{2 \pi}} \int_{-\pi }^{+\pi } \intpar{q} \Exp{-\imag k q} \ket{\psi_q^{(n)}} .
\end{align*}
For sufficiently deep lattices, the Wannier states with different $\ExcNum$ are localized around different lattice potential minima and are identical up to translation. We note that, within each energy band, the overlap between the quasi-momentum eigenstates and the Wannier states
\begin{align}
  \label{eq:WannierQuasiMomentumOverlap}
  \braket{\psi_q^{(n)} | w_k^{(m)}}
  &= \frac{\delta_{m,n}}{\sqrt{2\pi}} \Exp{- \imag k q} .
\end{align}
is similar to the usual overlap between position eigenstates and momentum eigenstates.

The Hamiltonian is block-diagonal in the basis of Wannier states, and the coupling of Wannier states at different locations is given by
\begin{align*}
  \braket{w_j^{(m)} | \Op H_\Latt | w^{(n)}_k} = -\delta_{m,n} J^{(n)}_{\abs{j - k}} ,
\end{align*}
where
\begin{align*}
  J^{(n)}_k
  &= -\frac{1}{2 \pi}  \int_{-\pi}^{+\pi} \intpar{q} \Exp{\imag k q} E_{q}^{(n)} .
\end{align*}
This shows that $J^{(n)}_k$ is the Fourier transform of the energy bands as a function of $q$ and the dispersion relations can be written as
\begin{align*}
  E_{q}^{(n)}
  &= -\sum_{k=0}^\infty J^{(n)}_k \Exp{-\imag k q}
  = - J^{(n)}_0 - 2 \sum_{k=1}^\infty J^{(n)}_k \cos(k q).
\end{align*}
For deep potentials the energy bands are relatively flat, and the higher order cosine terms are suppressed. This justifies the tight binding approximation in which one retains only the nearest lattice site coupling, and in the following we will suppress the band index $(n)$, and focus on the lowest band described by the tight binding Hamiltonian
\begin{align}
  \Op H_{\TB}
  &= - J_{1}\sum_{k = -\infty}^\infty   \tuborg{\ket{w_{k - 1}}  \bra{w_k} + \ket{w_{k + 1}}  \bra{w_k}}
  \nonumber  
  \\&=-2J_{1} \cos(\Op P).
\end{align}

\subsection{Harmonic confinement}

Adding a harmonic confinement to the lattice potential is adequately described by adding the term  $k \Op X^2$ with the spring constant $k$ to the Hamiltonian.
For deep lattice potentials, the Wannier state $\ket{w_j}$ is well localized at $X = j$, so we make the approximation
to replace $\Op X$ by the discrete quasi-position operator of the lowest band
\begin{align*}
  \Op W &= \sum_{j = - \infty}^{\infty} j  \ket{w_{j}} \bra{w_{j}} .
\end{align*}
Introducing a rescaling of the Hamiltonian by  $4J_1$ and defining $\kappa = k/ 4 J_1$  we end up with the Hamiltonian
\begin{align*}
  \Op H
  &= \frac{\Op H_\TB}{4 J_1} + \frac{k \Op W^2}{4 J_1}
  = \kappa \Op W^2 - \frac{\cos(\Op P)}{2}.
\end{align*}

Similar to the usual relationship between continuous position and momentum operators, the discrete position operator $\Op W$ acts as a differentiation in the continuous quasi-momentum representation
\begin{align*}
  \braket{\psi_q | \Op W | \alpha}
  &= \imag \pardd{}{q} \braket{\psi_q  | \alpha} .
\end{align*}
which is easily derived by inserting a resolution of the identity in Wannier states and using the overlap formula~(\ref{eq:WannierQuasiMomentumOverlap}). Therefore, we arrive at the quasi-momentum expression of the single particle Hamiltonian 
\begin{align}
  \braket{\psi_q | \Op H  | \alpha}
  &= \af{- \kappa \pardd[2]{}{q} -  \frac{\cos(q)}{2}} \braket{\psi_q | \alpha}
  \label{eq:SingleParticleHamiltonianInMomentumSpace}
\end{align}
At this point we make the curious observation \cite{RevModPhys.58.361,0305-4470-19-7-004} that, after having restricted the Hilbert space to the lowest energy band and having added a quasi-harmonic confinement, the Hamiltonian in momentum space~(\ref{eq:SingleParticleHamiltonianInMomentumSpace}) has the same form as the original optical lattice Hamiltonian~(\ref{eq:EnergyEigenequation}) in position space. In both cases, the Schrödinger equation takes the form of the Mathieu equation, but contrary to the case~(\ref{eq:EnergyEigenequation}) where we look for eigenstates with any quasi-momentum, here we will only look for \emph{periodic} eigenstates for~(\ref{eq:SingleParticleHamiltonianInMomentumSpace}), i.e. with zero ``quasi-position''.

\subsection{Interacting bosons}
In an ultra-cold gas of bosons, the interaction between the particles is adequately described by the two-particle contact interaction operator $\Op U_\Int$ with the matrix elements
\begin{align*}
  &\braket{X_1 ; X_2 | \Op U_\Int | X_3 ; X_4}
  \\& \qquad = g \delta(X_1 - X_3) \delta(X_2 - X_4) \delta(X_3 - X_4)
\end{align*}
for some interaction strength $g$. In the tight binding approximation, the Wannier states are localized at different lattice sites, and one may neglect matrix elements of the interaction potential with Wannier product states located on different sites. We thus end up with the following effective interaction operator acting on two-particle states
\begin{align}
  \Op U_\Int^\Eff
  &= G \sum_j \ket{w_j ; w_j} \bra{w_j ; w_j} ,
  \label{eq:EffectiveInteractionOperator}
\end{align}
where the strength parameter is given by
\begin{align*}
  G &= g \int \intpar{X} \abs{w_0(X)}^4 .
\end{align*}
A more rigorous treatment of the parameters of the Bose-Hubbard model can be found in e.g. \cite{PhysRevA.80.013404}.
Using the relation~(\ref{eq:WannierQuasiMomentumOverlap}), we can calculate the matrix elements of the effective interaction operator in quasi-momentum space
\begin{align*}
   \braket{\psi_{q_1} ; \psi_{q_2} | \Op U_\Int^\Eff | \psi_{q_3} ; \psi_{q_4}}
  &= \frac{G}{2\pi} \delta(q_3 + q_4 - q_1 - q_2) .
\end{align*}
which shows that the interaction conserves the total quasi-momentum and is independent of its value.


A system of two identical bosons in an optical lattice with harmonic confinement, which interact by the contact interaction is described by the Hamiltonian
\begin{align}
  \Op H
  &= \kappa (\Op W_1^2 + \Op W_2^2)
  - \frac{\cos(\Op P_1)}{2}
  - \frac{\cos(\Op P_2)}{2}
  + \Op U
  \label{eq:TwoParticleHamiltonianBeforeSeparation}
\end{align}
with $\Op U = \Op U_\Int^\Eff / 4 J_1$.

\section{Relative- and center-of-mass quasi-momenta}
In the quasi-momentum representation, the cosine terms of~(\ref{eq:TwoParticleHamiltonianBeforeSeparation}) can be written as
\begin{align*}
  &\frac{\cos(\Op P_1) + \cos(\Op P_2)}{2} \ket{\psi_{q_1} ; \psi_{q_2}}
  \\& \qquad= \frac{\cos(q_1) + \cos(q_2)}{2} \ket{\psi_{q_1} ; \psi_{q_2}}
  \\& \qquad= \cos\af{\frac{q_1 + q_2}{2}} \cos\af{\frac{q_2 - q_1}{2}}  \ket{\psi_{q_1} ; \psi_{q_2}}
  \\& \qquad\equiv \cos\af{\frac{\Op Q_+}{2}} \cos\af{\frac{\Op Q_-}{2}} \ket{\psi_{q_1} ; \psi_{q_2}},
\end{align*}
where we have defined new operators by their action on quasi-momentum eigenstates,
\begin{align*}
  \Exp{\imag \Op Q_\pm / 2}  \ket{\psi_{q_1} ; \psi_{q_2}}
  &\defi \Exp{\imag (q_2 \pm q_1)/ 2}  \ket{\psi_{q_1} ; \psi_{q_2}}.
\end{align*}

\begin{figure}
  \centering
  \includegraphics{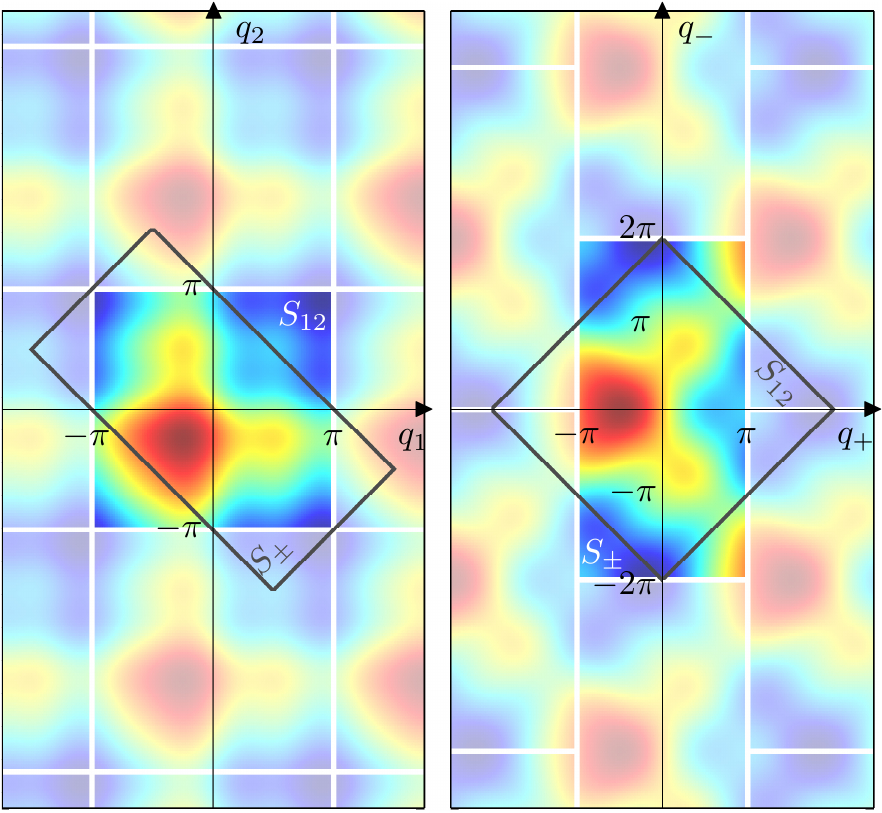}
  \caption{(color online) Quasi-momentum of the two particles vs. relative and center-of-mass quasi-momentum.
    Left: The first Brillouin zone $S_{12}$ in the $(q_1,q_2)$-plane is emphasized and repeated in each direction. The color coding indicates the values of a function that is periodic in both variables with period $2\pi$ and illustrates the required periodicity. The set $S_\pm$ which contains exactly one representative of each point from $S_{12}$ is shown by the gray rectangle.
  Right: The same function is shown but in the $(q_+,q_-)$-coordinate system. The set $S_\pm$ is emphasized and repeated, but with a different tiling than for $S_{12}$ in the left panel.}
  \label{fig:RelativeAndCOMQuasiMomentum}
\end{figure}

The introduction of these operators suggest to re-parameterize the quasi-momentum basis states $\ket{\psi_{q_1} ; \psi_{q_2}}$ in terms of their sum and difference:
\begin{align*}
  q_\pm = q_2 \pm q_1 .
\end{align*}
The quasi-momentum eigenstates states are defined for pairs of $q_1$ and $q_2$ in the set
\begin{align*}
  S_{12} &= ]-\pi ; \pi] \ \times \ ]-\pi ; \pi] ,
\end{align*}
corresponding to a diamond shaped area in the coordinate plane of $q_\pm$ as shown in figure~\ref{fig:RelativeAndCOMQuasiMomentum}. If we choose the values of $(q_+ , q_-)$ in the set
\begin{align*}
  S_\pm &= ]-\pi ; \pi] \ \times \ ]-2\pi ; 2\pi] ,
\end{align*}
then each point from $S_{12}$ is represented exactly once as is evident from figure~\ref{fig:RelativeAndCOMQuasiMomentum}. This means that we can re-parametrize the quasi-momentum eigenstates as
\begin{align*}
  \ket{q_+ , q_-}
  &= \frac{1}{\sqrt{2}} \ket{\psi_{(q_+ - q_-)/2} ; \psi_{(q_+ + q_-)/2}}
  \\
  \ket{\psi_{q_1} ; \psi_{q_2}}
  &= \sqrt{2} \ket{q_1 + q_2 , q_2 - q_1}
\end{align*}
where the front factor is chosen to preserve orthonormality, such that we have the resolution of identity
\begin{align}
  \Identity
  &= \int_{-\pi}^{+\pi} \intparNS{q_+} \int_{-2\pi}^{+2\pi} \intparNS{q_-} \ket{q_+ , q_-}\bra{q_+ , q_-}
\end{align}
The corresponding discrete relative and center-of-mass position operators
\begin{align*}
  \Op W_\pm &\defi \frac{\Op W_2 \pm \Op W_1}{2}
\end{align*}
act in the following way
\begin{align*}
  \braket{q_+ , q_- | \Op W_\pm | \alpha}
  &= \imag \pardd{}{q_\pm} \braket{q_+ , q_- | \alpha} .
\end{align*}
and the interaction operator $\Op U$ has the following representation in terms of the relative and center-of-mass quasi-momentum states
\begin{align*}
  \braket{q_+ , q_- | \Op U | \alpha}
  &= \gamma \int_{-2\pi}^{+2\pi} \intpar{q_-'}  \braket{q_+ , q_-' | \alpha}
\end{align*}
with $\gamma = G / (16 \pi J_1)$.

The two-atom Hamiltonian can now be written,
\begin{align}
  \Op H
  &= 2 \kappa (\Op W_+^2 + \Op W_-^2)
  - \cos\af{\frac{\Op Q_+}{2}} \cos\af{\frac{\Op Q_-}{2}}
  + \Op U .
  \label{eq:HamiltonianToSolve}
\end{align}
The Schrödinger equation with the Hamiltonian~(\ref{eq:HamiltonianToSolve}) can be solved accurately for a wide range of parameters (see Appendix~\ref{sec:SolvingNonapproximatedEquation}). The resulting eigenenergies and the wave functions~(\ref{eq:ExactSolutionWaveFunction}) will be used as reference for our analysis by the Born-Oppenheimer separation of the motional degrees of freedom which will be derived in the following section.

\section{Born-Oppenheimer separation}

\subsection{Derivation}
In order to separate the relative and the center-of-mass motion of the system, we write the Hamiltonian in~(\ref{eq:HamiltonianToSolve}) as
\begin{align}
  \Op H &= \Op H_- + 2 \kappa \Op W_+^2
  \label{eq:HamiltonianInTwoParts}
\end{align}
where $\Op H_-$ contains all operators dealing with the relative motion:
\begin{align*}
  \Op H_-
  &= 2 \kappa \Op W_-^2
  - \cos\af{\frac{\Op Q_+}{2}} \cos\af{\frac{\Op Q_-}{2}}
  + \Op U .
\end{align*}
We note that $\Exp{\imag \Op Q_+/2}$ commutes with $\Op H_-$ and we define their joint eigenstates $\ket{q_+ , \ExcNum}$:
\begin{align}
  \Op H_- \ket{q_+ , \ExcNum} &= \epsilon_\ExcNum(q_+)\ket{q_+ , \ExcNum} \label{eq:FirstBornOppenheimerEquation0}\\
  \Exp{\imag \Op Q_+/2} \ket{q_+ , \ExcNum} &= \Exp{\imag q_+/2} \ket{q_+ , \ExcNum}
\end{align}
with the following orthogonality relations
\begin{align}
  \braket{q_+ , \ExcNum | q_+' , \ExcNum'}
  &= \delta_{\ExcNum,\ExcNum'} \delta(q_+ - q_+') .
  \label{eq:qkOrthogonalityRelation}
\end{align}
The states $\ket{q_+ , \ExcNum}$ can be expanded
\begin{align}
  \ket{q_+ , \ExcNum} &= \int_{-2\pi}^{+2\pi} \intpar{q_-'} A_\ExcNum^{(q_+)}(q_-') \ket{q_+ , q_-'}.
  \label{eq:ExpansionOfEigenstatesOfFirstBOEquation}
\end{align}
and the orthogonality relation~(\ref{eq:qkOrthogonalityRelation}) implies
\begin{align*}
  \int_{-2\pi}^{+2\pi} \intpar{q_-} [A_\ExcNum^{(q_+)}(q_-)]^\CC A_{\ExcNum'}^{(q_+)}(q_-)
  &= \delta_{\ExcNum,\ExcNum'} .
\end{align*}

Any eigenstate of the full Hamiltonian~(\ref{eq:HamiltonianInTwoParts}) can be expanded as
\begin{align}
  \ket{\psi} &= \int_{-\pi}^{+\pi} \intpar{q_+'} \sum_\ExcNum C^{(\ExcNum)}(q_+') \ket{q_+' , \ExcNum}.
  \label{eq:expandeigen}
\end{align}
where the  expansion coefficients $C^{(\ExcNum)}(q_+')$ are found by applying the Hamiltonian~(\ref{eq:HamiltonianInTwoParts}) to the expanded wave function (\ref{eq:expandeigen}) and using~(\ref{eq:FirstBornOppenheimerEquation0})
\begin{align*}
  \Op H \ket{\psi}
  &= \int_{-\pi}^{+\pi} \intparNS{q_+'} \sum_\ExcNum C^{(\ExcNum)}(q_+') \tuborg{\epsilon_\ExcNum(q_+') + 2 \kappa \Op W_+^2}\ket{q_+' , \ExcNum} .
\end{align*}
In the $(q_+ , q_-)$-representation for the state vector, the eigenvalue equation takes the form of coupled differential equations
\begin{align}
  & E \sum_\ExcNum A_\ExcNum^{(q_+)}(q_-) C^{(\ExcNum)}(q_+)  \nonumber
  \\& \quad \quad = \sum_\ExcNum \tuborg{\epsilon_\ExcNum(q_+) -  2 \kappa \pardd[2]{}{q_+}} A_\ExcNum^{(q_+)}(q_-) C^{(\ExcNum)}(q_+)  .
  \label{eq:BeforeBornOppemheimerApproximation}
\end{align}

The goal of the following analysis is to find an approximation for the eigenstates, which is easier to apply numerically and which offers insights into their internal structure and dynamics. To this end, we assume that the states $\ket{q_+ , \ExcNum}$, described by $q_-$ wave functions $A_\ExcNum^{(q_+)}(q_-)$ depend only weakly on the argument $q_+$. Eliminating thus the partial derivatives of $A_\ExcNum^{(q_+)}(q_-)$ with respect to $q_+$ in the evaluation of the right hand side of~(\ref{eq:BeforeBornOppemheimerApproximation}), and using the orthogonality of the  $A_\ExcNum^{(q_+)}(q_-)$ functions, we arrive at the following approximate equation for the expansion coefficients
\begin{align}
  \epsilon_\ExcNum(q_+) C^{(\ExcNum)}(q_+)  - 2 \kappa \pardd[2]{C^{(\ExcNum)}(q_+)}{q_+}
  &= E C^{(\ExcNum)}(q_+) .
  \label{eq:SecondBornOppenheimerEquation1}
\end{align}
This has the form of a Schrödinger equation for a single particle in the potential $\epsilon_\ExcNum(q_+)$. For each energy potential we can find discrete eigenenergies $E_\SubExcNum^{(\ExcNum)}$ and associated eigenfunctions $C_\SubExcNum^{(\ExcNum)}$ that solve (\ref{eq:SecondBornOppenheimerEquation1}) and yield approximate eigenstates $\ket{\psi_\SubExcNum^{(\ExcNum)}}$ for the full Hamiltonian~(\ref{eq:HamiltonianInTwoParts})
\begin{align}
  \braket{q_+ , q_- | \psi_\SubExcNum^{(\ExcNum)} }
  &= C_\SubExcNum^{(\ExcNum)}(q_+) A_\ExcNum^{(q_+)}(q_-)
  \label{eq:ApproximateSolutionQuasiMomentumWaveFunction}
\end{align}

Note the formal similarity of this reduction of the problem with the use of the Born-Oppenheimer approximation in chemistry. In the latter, the wave function is expanded as a product of wave functions in nuclear and electronic coordinates, and due to the large difference in mass and hence in energy and time scales, the electronic wave functions are supposed to follow changes in the slow nuclear coordinates adiabatically. 

In our case, the two particles have identical masses, and in the absence of mutual interaction, the relative and center-of-mass motion occur on similar time scales, and the Born-Oppenheimer approximation should not be valid. But, as we increase the attractive interaction between the atoms, bound states are formed, and the relative position develops a new, faster time scale given by the binding energy.
 Our separation is carried out and motivated in the quasi-momentum picture, where a further observation may be in order: a strongly bound state in the relative position coordinate corresponds to a very extended wave function in the relative momentum, while the center-of-mass momentum may be well defined. This supports the assumption that the dominant contribution to the second derivative in~(\ref{eq:BeforeBornOppemheimerApproximation}) stems from the $q_+$ wave function $C_\SubExcNum^{(\ExcNum)}(q_+)$, and hence  that the derivative of  $A_\ExcNum^{(q_+)}(q_-)$ with respect to $q_+$ may be neglected.   

Since our approximate separation of the variables is mathematically equivalent to the usual Born-Oppenheimer approximation, albeit carried out in quasi-momentum representation rather than position representation, we will refer to is as ``the Born-Oppenheimer approximation'' in the following.

\subsection{Application}
\label{sec:Application}

Before we apply the Born-Oppenheimer approximation, let us consider how we expand states onto the center-of-mass and relative quasi-momentum eigenstates. Every state $\ket\phi$ can be expanded in both the two-particle quasi-momentum basis, and in the basis of relative and center-of-mass quasi-momenta.
\begin{align*}
  \ket\phi
  &=
  \begin{cases}
    \int\limits_{-\pi}^{+\pi} \intparNS{q_1} \int_{-\pi}^{+\pi} \intpar{q_2} \alpha(q_1 , q_1) \ket{\psi_{q_1} ; \psi_{q_2}}
    \\
    \int\limits_{-\pi}^{+\pi} \intparNS{q_+} \int_{-2\pi}^{+2\pi} \intpar{q_-} \beta(q_+ , q_-) \ket{q_+ , q_-} .
  \end{cases}
\end{align*}
While $\ket{q_1 , q_2}$ and $\ket{q_+ , q_-}$ are defined for $(q_1 , q_2) \in S_{12}$ and $(q_+ , q_-) \in S_\pm$, respectively, we can  look for functions defined on the entire $\mathbb R^2$ and restrict the solution afterwards. In this approach, the function $\alpha$ is periodic in both variables with period $2\pi$, and this enforces $\beta$ to obey the symmetry
\begin{align}
  \beta(q_+ + 2 \pi , q_- \pm 2 \pi ) = \beta(q_+ , q_-) 
  \label{eq:RequiredSymmetry}
\end{align}
c.f. the tiling of $\mathbb R^2$ with replicas of $S_\pm$ in the right panel of figure~\ref{fig:RelativeAndCOMQuasiMomentum}.
Thus, a necessary---but not sufficient---condition is that $\beta$ is periodic in both $q_+$ and $q_-$ with periodicity $4 \pi$. We are considering bosons and the state must be symmetric under the exchange of the two particles, $(q_+ , q_-) \mapsto (q_+ , -q_-)$, 
which implies the further constraint
\begin{align}
  \beta(q_+ , q_-) &= \beta(q_+ , -q_-) .
  \label{eq:SymmetrizationRequirement}
\end{align}

Using these arguments on~(\ref{eq:ApproximateSolutionQuasiMomentumWaveFunction}) we conclude that we are looking for solutions such that $A_\ExcNum^{(q_+)}(q_-)$ is even and periodic in $q_-$ with period $4\pi$, and such that the \emph{product} of $C_\SubExcNum^{(\ExcNum)}(q_+)$ and $A_\ExcNum^{(q_+)}(q_-)$ is periodic in $q_+$ with the same period. Furthermore, the product must satisfy the relation~(\ref{eq:RequiredSymmetry}).

\subsubsection{The first Born-Oppenheimer equation}

To apply the Born-Oppenheimer approximation, we must first find the eigenstates of $\Op H_-$ and their eigenenergies, and using the formal expansion of the states~(\ref{eq:ExpansionOfEigenstatesOfFirstBOEquation}), the eigenvalue equation~(\ref{eq:FirstBornOppenheimerEquation0}) leads to the equation
\begin{align}
  \epsilon_\ExcNum(q_+) A_\ExcNum^{(q_+)}(q_-)
  &= \kantpar{- 2\kappa \pardd[2]{}{q_-}
    -  F(q_+)\cos\af{\frac{q_-}{2}}
  }A_\ExcNum^{(q_+)}(q_-)
  \nonumber
  \\& \qquad \quad +
  \gamma \int_{-2\pi}^{+2\pi} \intpar{q_-'}  A_\ExcNum^{(q_+)}(q_-')
  \label{eq:FirstBornOppenheimerEquation1}
\end{align}
where $F(q_+) = \cos\af{\frac{q_+}{2}}$. For each value of $q_+$, this equation has the form of a Schrödinger equation with argument $q_-$, and with a periodic $\cos(\frac{q_-}{2})$ potential with amplitude $F(q_+)$ and a non-local potential with strength $\gamma$. Solutions which are periodic in $q_-$ with period $4 \pi$ are readily found by Fourier expansion of $A_\ExcNum^{(q_+)}(q_-)$ (see Appendix~\ref{sec:SolvingFirstBornOppenheimerEquation}), and these solutions can be chosen to be real-valued just like the zero quasi-momentum eigenstates for cosine potentials in position space.

The front factor $F(q_+)$ of the cosine potential is itself a cosine function of $q_+$ leading to two observations:
\begin{enumerate}
\item $F(q_+)$ is an even function of $q_+$ so Eq.~(\ref{eq:FirstBornOppenheimerEquation1}) is unaltered under the transformation $q_+ \mapsto -q_+$. Thus the solutions must be identical up to a complex factor, and since they are real-valued we can choose the solutions as
\begin{align}
  A_\ExcNum^{(q_+)}(q_-) &= A_\ExcNum^{(-q_+)}(q_-) .
  \label{eq:AkRelation1}
\end{align}
We could not have chosen a minus sign, since this would have made $A_\ExcNum^{(q_+)}$ vanish for $q_+ = 0$.

\item $F(q_+)$ changes to values of opposite sign when $q_+$ is increased by an amount of $2\pi$ and the cosine potential $\cos(q_-/2)$ in~(\ref{eq:FirstBornOppenheimerEquation1}) is effectively translated by half a period. For this translated potential the eigenvalues are the same, while the eigenfunctions are translated and scaled
  \begin{align}
    \epsilon_\ExcNum(q_+) &= \epsilon_\ExcNum(q_+ + 2\pi)
    \label{eq:EpsilonkPeriodic}
    \\
    A_\ExcNum^{(q_+)} (q_-) &= \xi_\ExcNum A_\ExcNum^{(q_+ + 2\pi)}(q_- \pm 2\pi) .
    \label{eq:AkRelation2}
  \end{align}
  The factor $\xi_\ExcNum$ may take the values $\pm 1$ since $A_\ExcNum^{(q_+)} (q_-)$ is real-valued for all values of $q_+$ and $q_-$.
\end{enumerate}
Applying the relations~(\ref{eq:AkRelation1}) and~(\ref{eq:AkRelation2}) for $q_+ = -\pi$ we get the relation
\begin{align}
  A_\ExcNum^{(+\pi)}(q_-)
  &= \xi_\ExcNum A_\ExcNum^{(+ \pi)}(q_- \pm 2\pi)
\end{align}
so we can determine $\xi_\ExcNum$ from the translational symmetries of $A_\ExcNum^{(+\pi)}$.

\begin{figure*}
  \centering
  \includegraphics{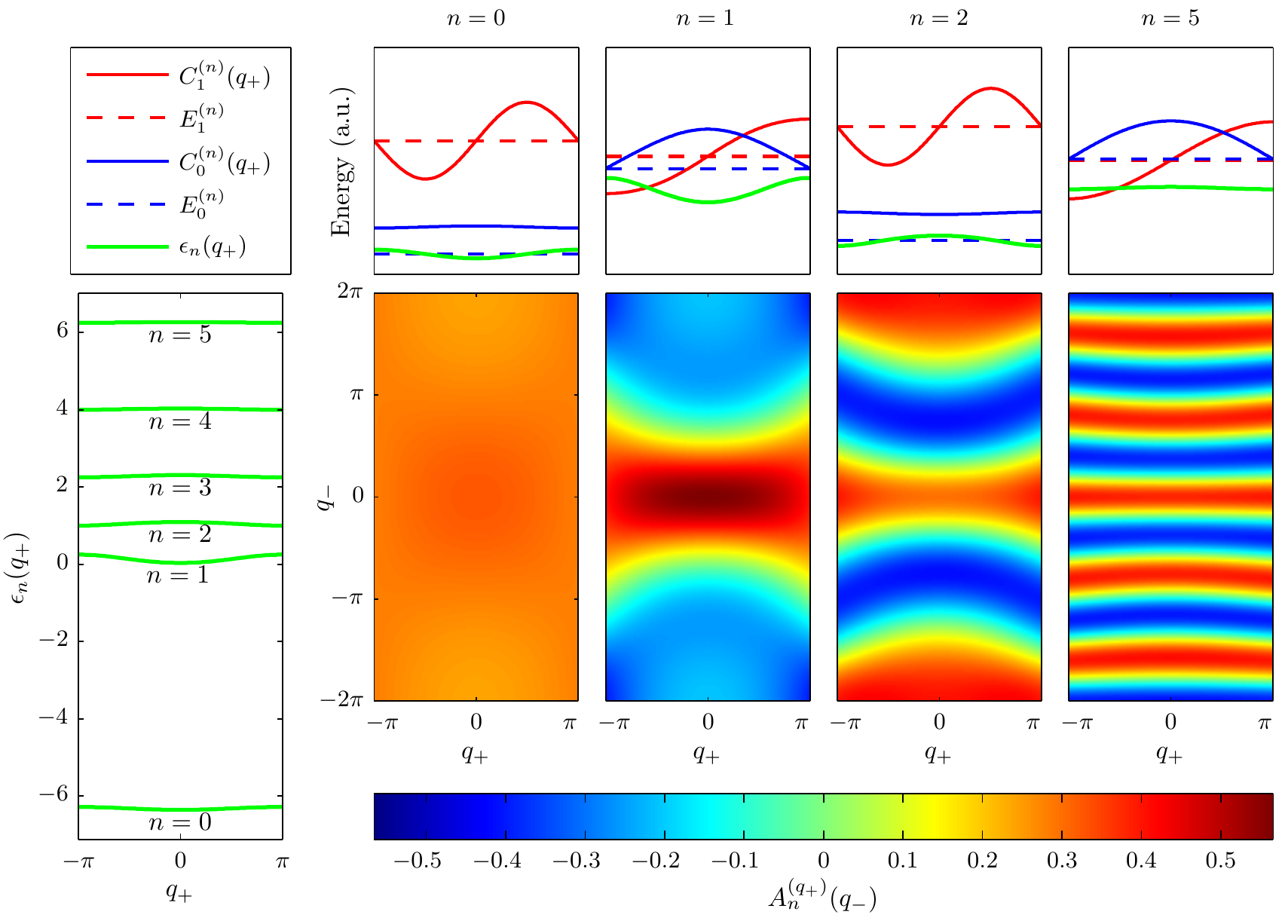}
  \caption{(color online) Energies and eigenfunctions found by solving the two Born-Oppenheimer equations for $\kappa = 0.5$ and $\gamma = -0.5$.
    Left panel: The six lowest potential curves $\epsilon_\ExcNum(q_+)$ found from the first Born-Oppenheimer equation.
    Upper panels: Magnification of four of the potential curves in the left panel.
    Lower panels: Eigenfunctions $A_\ExcNum^{(q_+)}(q_-)$ for the first Born-Oppenheimer equation shown for all values of $q_+$ for the corresponding $\ExcNum$-values.
    In the upper panels are shown (horizontal dashed blue/red lines) the two lowest energies $E_\SubExcNum^{(\ExcNum)}$ for $\SubExcNum = 0,1$ found from solving the second Born-Oppenheimer equation in the potential $\epsilon_\ExcNum(q_+)$ and the corresponding wave functions (solid blue/red lines).}
  \label{fig:ResultsOfBorn-OppenheimerEquations}
\end{figure*}

\subsubsection{The second Born-Oppenheimer equation}

Solving Eq.~(\ref{eq:FirstBornOppenheimerEquation1}) yields the potential $\epsilon_\ExcNum(q_+)$ which is periodic with period $2\pi$, and we are looking for functions $C_\SubExcNum^{(\ExcNum)}(q_+)$ that are periodic in $q_+$ with period $4\pi$. Therefore, Bloch's theorem tells us that we can choose a complete set of solutions as
\begin{align}
  C_\SubExcNum^{(\ExcNum)}(q_+)
  &= \Exp{\imag \delta_\ExcNum q_+/2} D_\SubExcNum^{(\ExcNum)}(q_+)
  \label{eq:BlochsTheoremOnSecondBornOppenheimerEquation}
\end{align}
where $D_\SubExcNum^{(\ExcNum)}$ is periodic with periodic $2 \pi$, and $\delta_\ExcNum = 0,1$. For $\delta_\ExcNum = 0$ the solution $C_\SubExcNum^{(\ExcNum)}(q_+)$ is thus periodic with period $2\pi$, whereas for $\delta_\ExcNum = 1$, it is antiperiodic. We require that the product of
$C_\SubExcNum^{(\ExcNum)}(q_+)$ and $A_\ExcNum^{(q_+)}(q_-)$
satisfies the symmetry~(\ref{eq:RequiredSymmetry}), and if we combine this with~(\ref{eq:AkRelation2}), we get the relation
\begin{align*}
  C_\SubExcNum^{(\ExcNum)}(q_+) A_\ExcNum^{(q_+)}(q_-)
  &= \xi_\ExcNum C_\SubExcNum^{(\ExcNum)}(q_+ + 2 \pi) A_\ExcNum^{(q_+)}(q_-)
\end{align*}
from which we conclude that $C_\SubExcNum^{(\ExcNum)}(q_+)$ must fulfill the symmetry
\begin{align*}
  C_\SubExcNum^{(\ExcNum)}(q_+ + 2 \pi) &= \xi_\ExcNum C_\SubExcNum^{(\ExcNum)}(q_+) .
\end{align*}
Comparing to~(\ref{eq:BlochsTheoremOnSecondBornOppenheimerEquation}) we see that  for $\xi_\ExcNum = -1$ we must choose $\delta_\ExcNum = 1$ and for $\xi_\ExcNum = +1$, we must use $\delta = 0$. We can solve~(\ref{eq:SecondBornOppenheimerEquation1}) by Fourier expansions of $D_\SubExcNum^{(\ExcNum)}$ and $\epsilon_\ExcNum$  (see Appendix~\ref{sec:SolvingSecondBornOppenheimerEquation}).

\begin{figure*}
  \centering
  \includegraphics{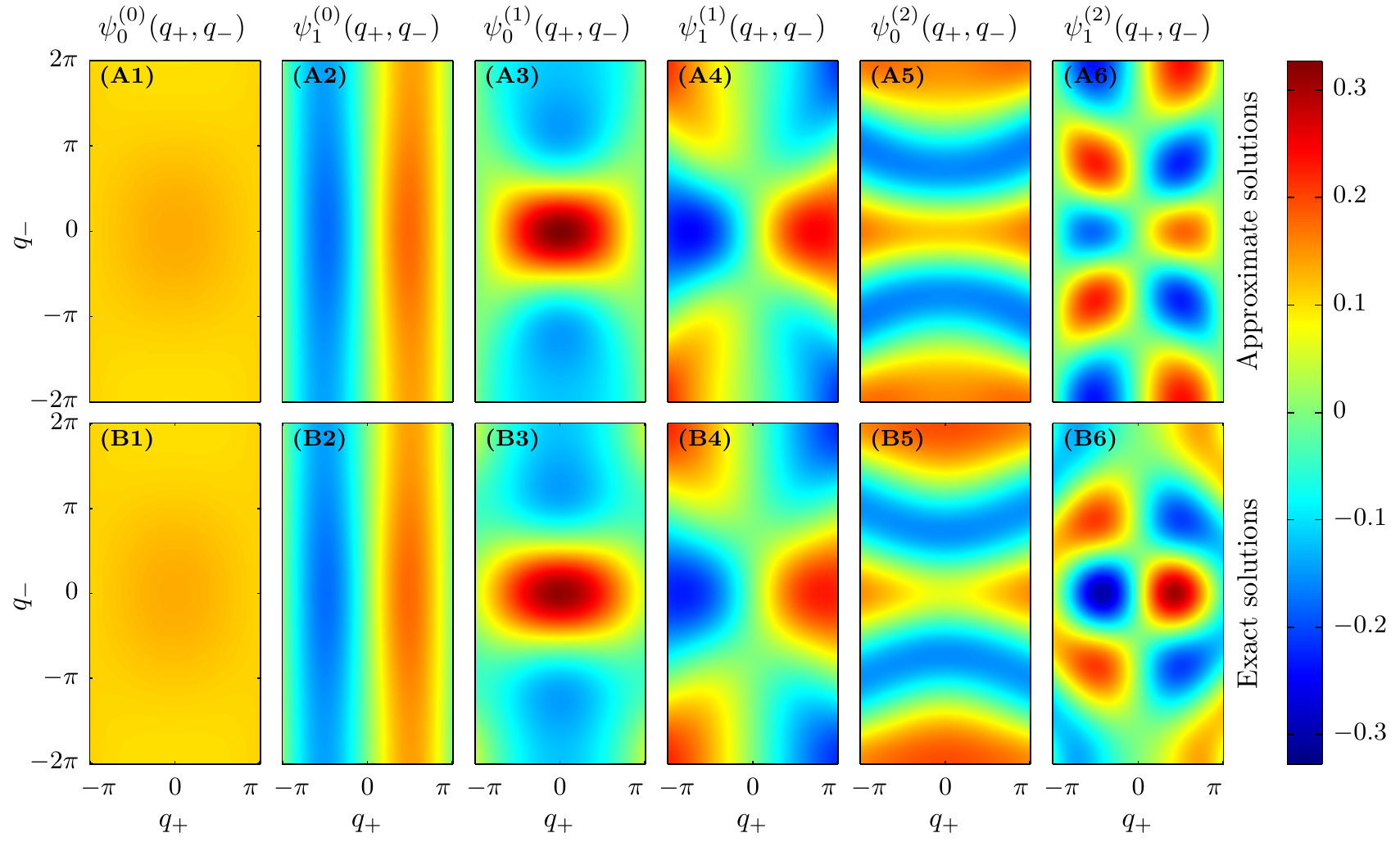}
  \caption{(color online) Upper panels: Quasi-momentum wave functions found by the Born-Oppenheimer approximation for $\kappa = 0.5$ and $\gamma = -0.5$. Columns 1--2 show the two lowest eigenstates in the lowest potential curve $\epsilon_0(q_+)$, corresponding to the $\ExcNum=0$ column in figure~\ref{fig:ResultsOfBorn-OppenheimerEquations}. Columns 3--4 correspond to the $\ExcNum=1$ column in figure~\ref{fig:ResultsOfBorn-OppenheimerEquations} and columns 5--6 correspond to the $\ExcNum = 2$ column in figure~\ref{fig:ResultsOfBorn-OppenheimerEquations}.
  Lower panels: The corresponding exact solutions.}
  \label{eq:ApproximateVSExactWaveFunction}
\end{figure*}

\section{Born-Oppenheimer solutions}

\subsection{Wave functions}

When solving the first Born-Oppenheimer equation~(\ref{eq:FirstBornOppenheimerEquation1}) we find eigenvalues $\epsilon_\ExcNum(q_+)$ and eigenfunctions $A_\ExcNum^{(q_+)}(q_-)$ for each value of $q_+ \in ]-\pi , +\pi]$. In the leftmost panel in figure~\ref{fig:ResultsOfBorn-OppenheimerEquations} is shown the six lowest potential curves $\epsilon_\ExcNum(q_+)$. The lowest potential curve is well separated from the higher ones which lie closer. Each of the potential curves has an energy variation which is typically small compared to the energy distance between the bands, and in the upper panels, a magnified view of the curves are shown. In the lower panels, eigenfunctions $A_\ExcNum^{(q_+)}(q_-)$ of the first Born-Oppenheimer equation are shown for four different values of $\ExcNum$.

The second Born-Oppenheimer equation uses the energies $\epsilon_\ExcNum(q_+)$ as potential functions in a Schrödinger like equation, and each of the upper panels in Figure~\ref{fig:ResultsOfBorn-OppenheimerEquations} shows the energy levels of the two lowest eigenstates ($\SubExcNum = 0,1$) in these potentials along with their eigenfunctions $C_\SubExcNum^{(\ExcNum)}(q_+)$. As we saw in the previous section, the function $C_\SubExcNum^{(\ExcNum)}(q_+)$ should be chosen periodic or anti-periodic depending on the value of $\xi_\ExcNum$. By studying the behavior of $A_\ExcNum^{(q_+)}(q_-)$ at $q_+ = \pm \pi$ one can see if $\xi_\ExcNum$ is $+1$ or $-1$ depending on whether the wave function $A_\ExcNum^{(\pm \pi)}(q_-)$ changes sign when translated by $\pi$ or not. For $\ExcNum = 0,2$ the solutions $C_\SubExcNum^{(\ExcNum)}(q_+)$ to the second Born-Oppenheimer equation must be periodic with period $2\pi$, while for $\ExcNum = 1,5$ the solutions $C_\SubExcNum^{(\ExcNum)}(q_+)$ must be chosen \emph{anti}periodic.

Total Born-Oppenheimer solutions to the two-atom Hamiltonian are shown in Fig.~\ref{eq:ApproximateVSExactWaveFunction}, where
panels (A1--2) correspond to the approximate solutions from the $\ExcNum = 0$ case of figure~\ref{fig:ResultsOfBorn-OppenheimerEquations}, panels (A3--4) correspond to the $\ExcNum = 1$ case, and panels (A5--6) correspond to the $\ExcNum = 2$ case. In the lower panels of figure~\ref{eq:ApproximateVSExactWaveFunction}, the corresponding exact two-atom eigenstates are shown. There is a good  agreement between the exact and approximate solutions, especially for the low excitations of the lowest bands.

\subsection{Energies}

\begin{figure}[tb]
  \centering
  \includegraphics[width=\columnwidth]{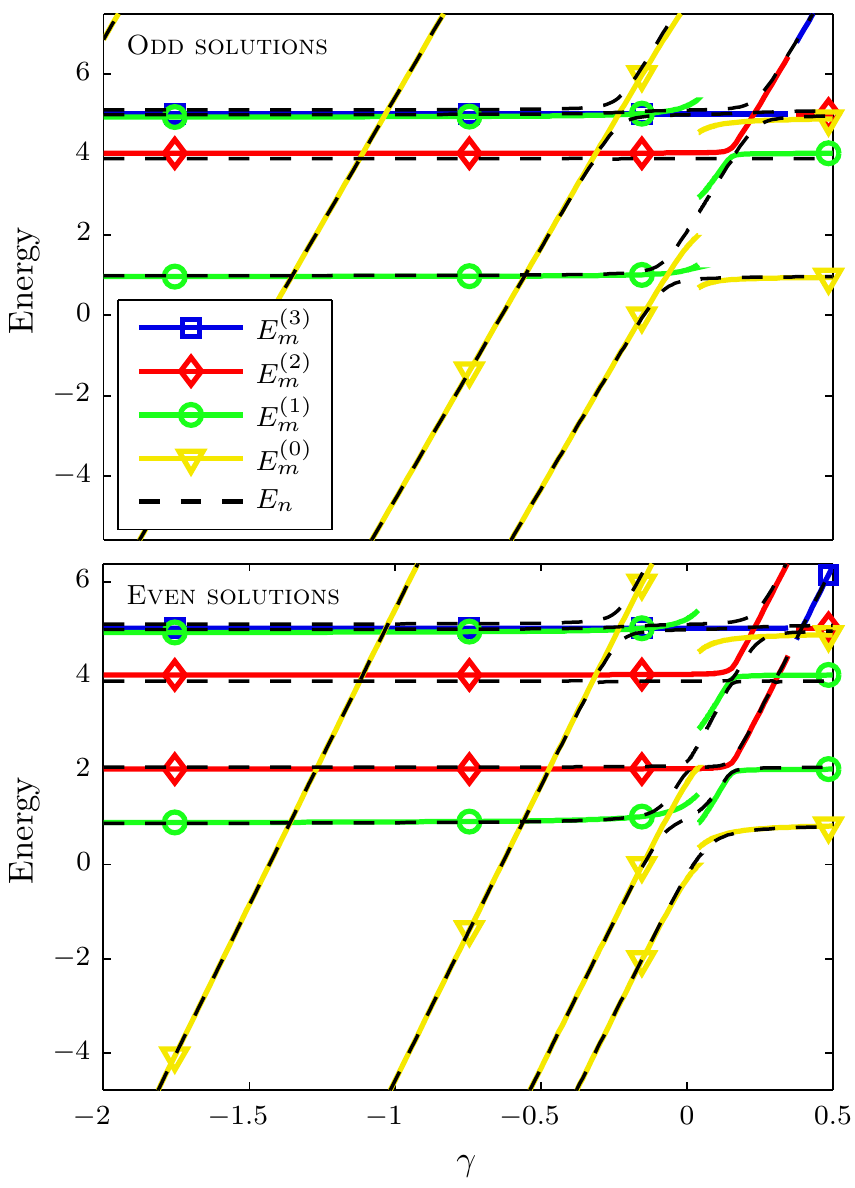}
  \caption{(color online) Exact and approximate energies as a function of $\gamma$ for $\kappa=1$. Exact energies are plotted by the black dashed lines, and for $\ExcNum = 0,1,2,3$ the energies $E_j^{(\ExcNum)}$ found from the Born-Oppenheimer equation are plotted in four different colors.}
  \label{fig:EnergiesGamma}
\end{figure}

In figure~\ref{fig:EnergiesGamma} both the exact and the approximative energies are shown for fixed $\kappa$ as functions of the scaled interaction strength $\gamma$. Except in the region where $\gamma$ is numerically small, there is reasonable agreement between the exact and the approximated energy levels. For negative $\gamma$ there is a clear grouping of the energy levels in two groups: Those that are nearly constant as a function of $\gamma$ and those that depend linearly on $\gamma$. Comparing to the approximate energies found by the Born-Oppenheimer approximation we see that the linear dependence comes from the fact that the position of the lowest potential curve varies linearly with $\gamma$, as we will see in the following.


\subsection{Approximate solution of the first Born-Oppenheimer equation}

To understand the behavior of the energy spectra, we start by analyzing the system in the limit where at least one of the two coefficients $\gamma$, $\kappa$ is (numerically) much larger than unity, so that we can find analytical approximations. This limit enables us to treat the term
\begin{align}
  - F(q_+) \cos\af{\frac{q_-}{2}}
  \label{eq:PertubationTerm}
\end{align}
in the first Born-Oppenheimer equation~(\ref{eq:FirstBornOppenheimerEquation1}) as a perturbation. When we neglect this term, we can choose a complete set of eigenfunctions as plane waves with wave number $\WaveNum/2$ for $\WaveNum = 0 , \pm 1 , \pm 2, \ldots$ and with energies
\begin{align}
  \tilde \epsilon_\WaveNum &= \frac{\kappa}{2} \WaveNum^2 +  4 \pi \gamma \delta_{\WaveNum,0} ,
  \label{eq:UnperturbedEnergies}
\end{align}
but only the even linear combinations are physically relevant. Note that the $\WaveNum$ does in general not coincide with the excitation number $\ExcNum$ as used in the first Born-Oppenheimer equation, where the energy curves $\epsilon_\ExcNum$ were sorted by energy.

The term $4 \pi \gamma$ contributes only for $\WaveNum = 0$ since all other plane waves integrate to zero in the second line of Eq.~(\ref{eq:FirstBornOppenheimerEquation1}). Even when the omission of~(\ref{eq:PertubationTerm}) is not valid, the integral still becomes substantial if $A_\ExcNum^{(q_+)}(q_-)$ has no nodes, whereas it is suppressed when there are sign changes in $A_\ExcNum^{(q_+)}(q_-)$.

In figure~\ref{fig:EnergiesGamma} we notice some discontinuities in the approximate energies, which can be explained in the following way.
Due to the linear dependence of the energy for the $\WaveNum = 0$ plane wave, its energy becomes degenerate with the higher levels, when $\gamma$ varies. More precisely, $\tilde \epsilon_0$ will cross $\tilde \epsilon_\WaveNum$ at the $\gamma$-value
\begin{align}
  \gamma_\WaveNum &= \frac{\kappa \WaveNum^2}{8\pi} .
  \label{eq:TildeEpsilonCrossing}
\end{align}
Without the symmetry requirement~(\ref{eq:RequiredSymmetry}) we could find two families of solutions  to the second Born-Oppenheimer equation for each potential curve $\epsilon_\ExcNum(q_+)$. Depending on the symmetries of the solution $A_\ExcNum^{(q_+)}(q_-)$ discussed in Sec.~\ref{sec:Application} we can only choose one of these families, and at each side of the energy crossing~(\ref{eq:TildeEpsilonCrossing}), we must discard one or the other and thus obtain a discontinuous energy dependence.

Now, we turn to the term~(\ref{eq:PertubationTerm}), the effect of which we will approximate using non-degenerate perturbation theory. Due to the orthogonality between the cosine functions, there are no first-order corrections. The second order corrections, on the other hand, give contributions of the form
\begin{align*}
  \Delta \tilde\epsilon_\WaveNum(q_+)
  &= a_\WaveNum F(q_+)^2
\end{align*}
where the amplitude $a_\WaveNum$ can be calculated (see Appendix~\ref{sec:PerturbativeTerms})
\begin{align}
  a_\WaveNum
  &=
  \begin{cases}
    - \frac{1}{\kappa - 8 \pi \gamma} & \WaveNum = 0 , \\
    \frac{1}{2\kappa - 16 \pi \gamma} -\frac{1}{6\kappa} & \WaveNum= \pm 1 , \\
    \frac{1}{\kappa (4\WaveNum^2 - 1)} & \text{otherwise}  .
  \end{cases}
  \label{eq:CorrectionAmplitudes}
\end{align}
This gives the perturbative approximation to the potential curves
\begin{align*}
  \tilde \epsilon_\WaveNum(q_+) +  \Delta \tilde\epsilon_\WaveNum(q_+)
  &=  \frac{\kappa}{2} \WaveNum^2 +  4 \pi \gamma \delta_{\WaveNum,0} + \frac{a_\WaveNum}{2}\af{ 1 +  \cos(q_+)}
\end{align*}
Due to the term $4 \pi \gamma$ in the expression for $\tilde\epsilon_0(q_+)$, the energies of the eigenfunctions in this potential change linearly with $\gamma$. For $\WaveNum \geq 1$ the position of $\tilde\epsilon_\WaveNum(q_+)$ depends less strongly on $\gamma$ and the eigenstates in these potentials have almost constant energy.

\subsection{Approximate solution of the second Born-Oppenheimer equation}

To analyze in more detail how the eigenenergies $E_\SubExcNum^{(\ExcNum)}$ are distributed we must take a closer look at the second Born-Oppenheimer equation which has the form of a Schrödinger equation for a particle of mass $\hslash^2 / 4 \kappa$ in the potential $\epsilon_\ExcNum(q_+)$. When the above perturbative treatment is valid, this potential is a cosine with amplitude $\abs{a_\WaveNum}/2$, so in order to estimate the eigenstates and energies, we must compare $\kappa$ and $\abs{a_\WaveNum}$. In the limit where we can neglect the $q_+$-dependence of $\tilde\epsilon_\WaveNum(q_+)$, the solutions can be well approximated by plane waves
$\Exp{\imag \SubExcNum q_+} / \sqrt{2\pi}$
with ``box potential''-energies
\begin{align*}
  \tilde E_\SubExcNum^{(\WaveNum)} &= \frac{\kappa}{2} \WaveNum^2 +  4 \pi \gamma \delta_{\WaveNum,0} +  2 \kappa \SubExcNum^2
\end{align*}
which depend quadratically on $\SubExcNum$. In the opposite limit where $\tilde\epsilon_\WaveNum(q_+)$ is a deep potential in~(\ref{eq:SecondBornOppenheimerEquation1}), we can approximate the cosine potential by a quadratic expansion around its minimum. The resulting equation is a Schrödinger equation for a particle in a harmonic oscillator of frequency
\begin{align*}
  \omega_\WaveNum = \frac{1}{\hslash} \sqrt{2 \kappa \abs{a_\WaveNum}} .
\end{align*}
For the lower part of the energy spectrum, the solutions are then well approximated by the usual harmonic oscillator eigenstate wave functions and the energies are equidistantly spaced with spacing $\hslash \omega_\WaveNum$:
\begin{align*}
  \tilde E_\SubExcNum^{(\WaveNum)} &= \af{\frac{\kappa}{2} \WaveNum^2 +  4 \pi \gamma \delta_{\WaveNum,0} + \frac{a_\WaveNum}{2}}  +  \af{\SubExcNum + \frac{1}{2}} \sqrt{2 \kappa \abs{a_\WaveNum}} .
\end{align*}

\begin{figure}[tb]
  \centering
  \includegraphics[width=\columnwidth]{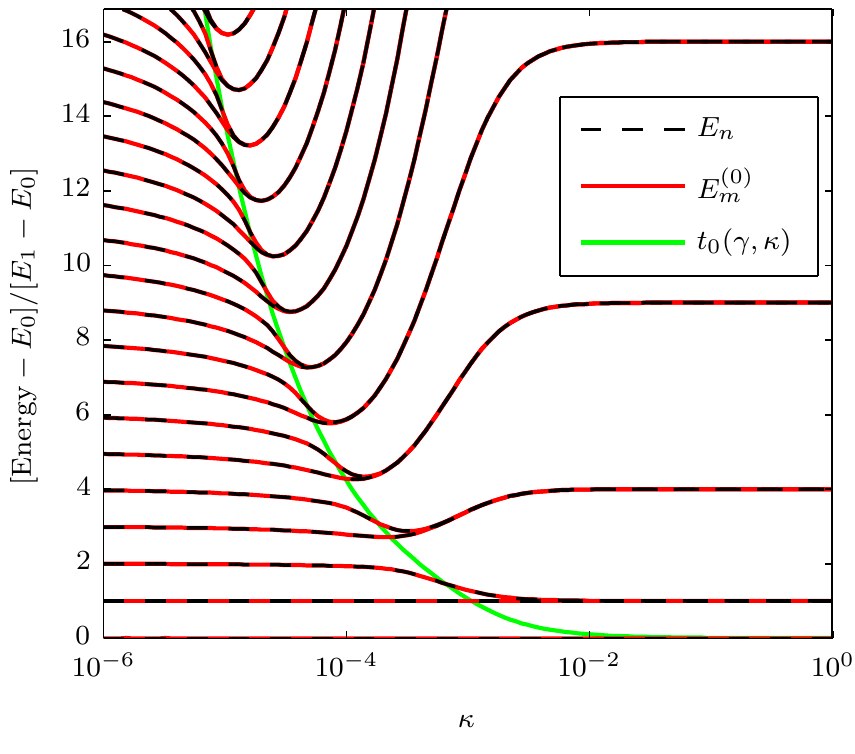}
  \caption{(color online) Exact and approximate energies as a function of $\kappa$ for $\gamma = -10$. The black dashed curves show the exact energies $E_n$, while the solid red curves show the approximate energies $E_\SubExcNum^{(0)}$ found from the lowest potential-curve in the second Born-Oppenheimer equation. The green curve shows the position of the maximum of the lowest potential curve $\epsilon_0(q_+)$ within the perturbative approximation. The exact ground state energy $E_0$, which varies with $\kappa$, has been subtracted from all energies, and afterward, the energies is scaled by the energy difference between the two lowest exact energy levels $E_1 - E_0$.}
  \label{eq:EnergyVsKappa}
\end{figure}

Figure~\ref{eq:EnergyVsKappa} illustrates the transition between the ``particle in a box'' and the ``harmonic oscillator'' regimes
by showing the exact and approximate energies $E_\SubExcNum^{(0)}$ as functions of $\kappa$ for fixed negative $\gamma$. Since the harmonic oscillator approximation is valid when the potential in~(\ref{eq:SecondBornOppenheimerEquation1}) is deep, it requires that $\kappa \ll \abs{a_0} = \abs{\kappa - 8 \pi \gamma}^{-1}$, so to capture the whole transition, the $\kappa$-axis is logarithmic. The energies are plotted after subtracting the ground state energy $E_0$ and scaling by the energy difference $E_1 - E_0$ between the first excited state and the ground state. For $\kappa \ll 1$ the harmonic oscillator spectrum is then revealed as levels with unit spacing. For $\kappa \to 1$, on the other hand, the curves become constant at $1 , 4 , 9, \ldots$ showing the quadratic dependence on $\SubExcNum$ (see also \cite{PhysRevA.72.033616}). We note that there is a perfect agreement between the exact and approximate energies shown in the figure. In the transition from the harmonic oscillator regime to the ``particle in a box'' regime, the energy levels group in pairs, which have the following explanation: For a deep potential curve $\epsilon_\ExcNum$ there is a significant energy difference between the first excited even and odd states, but when the potential curve is nearly constant, then even and odd solutions with a given wave number has almost the same energy.

No matter how deep the potential curve $\tilde\epsilon_\WaveNum(q_+)$ is, the harmonic approximation is not perfect, and above some energy the spectrum is ill-described by a harmonic oscillator spectrum. A simple estimate suggests that the description is good for eigenstates whose energies lie below the maximum of the potential curve, which is approximated by the unperturbed energies~(\ref{eq:UnperturbedEnergies}) plus a term depending on the sign of $a_\WaveNum$
\begin{align*}
  t_\WaveNum(\gamma,\kappa) &= \frac{\kappa}{2} \WaveNum^2 + 4 \pi \gamma \delta_{\WaveNum,0} + \frac{a_\WaveNum + \abs{a_\WaveNum}}{2} .
\end{align*}
In figure~\ref{eq:EnergyVsKappa} this (solid green) curve is shown for $\WaveNum = 0$ and agrees systematically with the border where the harmonic oscillator energy spectrum is significantly altered.

\section{Conclusion}

In the present paper we have considered two identical bosons on an infinite, discrete lattice with an additional harmonic confinement. In the tight binding approximation, the single particle physics in terms of quasi-momenta is described by the same equation as a single particle in a continuous cosine potential---namely the Mathieu equation. Adding a contact interaction yields a Hamiltonian which does not separate in relative and center-of-mass coordinates, even though the two-body interaction problem separates in both a homogeneous discrete lattice Hamiltonian and in a continuous harmonic oscillator.

By formulating the problem in quasi-momentum representation we can make an approximation which is mathematically equivalent to the usual Born-Oppenheimer approximation performed in position space in molecular physics: We thus find approximate solutions by first solving an equation for the relative quasi-momentum wave function that depends parametrically on the center-of-mass quasi-momentum. This yields potential curves for a Schrödinger equation for the center-of-mass coordinate, which is readily solved. Contrary to the usual Born-Oppenheimer approximation used to separate slow nuclear and fast electronic motion in molecules, in our system we have a tunable adiabaticity parameter, namely the strength of the inter-particle interaction.

In the solution of both the first and second Born-Oppenheimer equations we can identify the excitation degrees of freedom in the system. This provides physically motivated quantum numbers valid also for the exact eigenstates together with rules for which quantum numbers are allowed by symmetry considerations.

Finally, from the good agreement between the exact and approximate solutions we conclude that the Born-Oppenheimer approximation is well justified when the  energy scales for the relative and the center-of-mass motion of the two-particle quantum state are well-separated. We imagine that a similar separation may be useful for approximate first principle calculations on many other cold atom systems, e.g., with more particles and possibly with mixtures of different species.

\begin{acknowledgments}
The authors gratefully acknowledge discussions with Manuel Valiente.
\end{acknowledgments}

\appendix

\section{Solving the equations numerically}
\label{sec:SolvingEqutionsNumerically}

\subsection{Solving the non-approximated equation}
\label{sec:SolvingNonapproximatedEquation}
To solve the two-atom Schrödinger equation in the tight binding approximation we expand the state as
\begin{align}
  \ket{\alpha}
  &= \sum_{j ,k} F_{jk} \ket{w_j ; w_k} .
  \label{eq:ExpansionOfExactEigenstate}
\end{align}
The stationary Schrödinger equation with the Hamiltonian~(\ref{eq:HamiltonianToSolve}) yields the equation for the expansion coefficients
\begin{align*}
  E F_{j,k}
  &= 4 \pi \gamma \delta_{j,k} F_{j,k}
  +\kappa(j^2 + k^2) F_{jk}
  \\& \qquad - \frac{F_{j-1,k} + F_{j+1,k} + F_{j,k-1} + F_{j,k+1}}{4} .
\end{align*}
The original Hamiltonian is invariant under parity inversion of both particles so we can find a complete set of solutions of even and odd wave-functions. In terms of the expansion~(\ref{eq:ExpansionOfExactEigenstate}) this means that we can find solutions where
\begin{align*}
  F_{-j , -k} &= \Oddity F_{j,k}
\end{align*}
where $\Oddity$ can assume the values $\pm 1$. In addition, since we are dealing with two identical bosons, only symmetrized wave functions are physically meaningful, with implies that we have the symmetry
\begin{align*}
  F_{j,k} = F_{k,j} .
\end{align*}
For numerical purposes we enforce these requirements by hand in the following way. Instead of looking at all pairs $(j,k) \in \Integers^2$, we restrict our attention to those in the subset
\begin{align*}
  T
  &= \tuborg{(j,k) \in \Integers^2 \ \vert \ \abs{k} \leq j \leq j_\Max}
  \subset \Integers^2 .
\end{align*}
for some manually chosen $j_\Max$.
Using the symmetries we reformulate the recurrence equation such that it only involves coefficients from $T$.
The equation can be expressed as a matrix eigenvalue equation which is amenable to standard numerical diagonalization routines.
When all coefficients have been found---and properly normalized---the wave function in relative and center-of-mass quasi-momenta is given by
\begin{align}
  \braket{q_+ , q_- | \alpha}
  &= \sum_{j ,k} F_{jk} \braket{q_+ , q_- | w_j ; w_k}
  \nonumber
  \\&= \frac{1}{\sqrt{2} (2\pi)}\sum_{j ,k} F_{jk} \Exp{- \imag (j + k)q_+/2} \Exp{- \imag (k-j) q_-/2}
  \label{eq:ExactSolutionWaveFunction}
\end{align}

\subsection{Solving the first Born-Oppenheimer equation}
\label{sec:SolvingFirstBornOppenheimerEquation}
The solutions of~(\ref{eq:FirstBornOppenheimerEquation1}) are functions $A_\ExcNum^{(q_+)}(q_-)$ which are periodic in $q_-$ with period $4\pi$. Therefore, for each value of $q_+$ we make the expansion
\begin{align}
  A_\ExcNum^{(q_+)}(q_-)
  &= \frac{1}{\sqrt{4 \pi}}  \sum_j \alpha_{j,\ExcNum}^{(q_+)} \Exp{\imag j q_- / 2} ,
  \label{eq:FourierExpansionOfAn}
\end{align}
and obtain the tridiagonal recurrence relation,
\begin{align}
  &\frac{ F(q_+) }{2} \kantpar{\alpha_{j-1,\ExcNum}^{(q_+)} + \alpha_{j+1,\ExcNum}^{(q_+)}}
  \nonumber
  \\& \qquad \qquad  =
  \af{\frac{\kappa}{2}j^2 + 4 \pi \gamma \delta_{j,0} - \epsilon_\ExcNum(q_+) } \alpha_{j,\ExcNum}^{(q_+)} .
  \label{eq:FirstBornOppenheimerEquation2}
\end{align}
To accommodate the bosonic nature of the particles, we only look for even solutions to~(\ref{eq:FirstBornOppenheimerEquation1}), so we only need to consider  terms $\alpha_{j,\ExcNum}^{(q_+)}$ with $j \geq 0$, and for $j=0$ we use
\begin{align}
  F(q_+) \alpha_{1,\ExcNum}^{(q_+)}
  &= \kantpar{ 4 \pi \gamma - \epsilon_\ExcNum(q_+) } \alpha_{0,\ExcNum}^{(q_+)} .
\end{align}
By expressing the recurrence relation as a matrix eigenvalue equation, this can be truncated and solved with good accuracy.

\subsection{Solving the second Born-Oppenheimer equation}
\label{sec:SolvingSecondBornOppenheimerEquation}
We solve the second Born-Oppenheimer equation using the results from the first Born-Oppenheimer equation. First, coefficients in the expansion
\begin{align}
  \epsilon_\ExcNum(q_+) = \sum_k \beta^\ExcNum_k \Exp{\imag j q_+}
  \label{eq:FourierExpansionOfEpsilonk}
\end{align}
are determined by a discrete Fourier transformation. We then use the expansion 
\begin{align*}
  C_\SubExcNum^{(\ExcNum)}(q_+)
  &= \frac{1}{\sqrt{2\pi}} \sum_l \gamma_{l}^{\SubExcNum,\ExcNum} \Exp{\imag (l + \frac{\delta_\ExcNum}{2})q_+} .
\end{align*}
in~(\ref{eq:SecondBornOppenheimerEquation1}) together with the expansion~(\ref{eq:FourierExpansionOfEpsilonk}) which yields the following equation for the $\gamma$-coefficients
\begin{align}
  \sum_k \beta_{k}^\ExcNum \gamma_{l-k}^{\SubExcNum,\ExcNum} +  2\kappa \af{l +  \frac{\delta_\ExcNum}{2}}^2 \gamma_{l}^{\SubExcNum,\ExcNum}
  &= E_\SubExcNum^{(\ExcNum)} \gamma_{l}^{\SubExcNum,\ExcNum}  .
  \label{eq:SecondBornOppenheimerEquation2}
\end{align}
Since the potential energy curves $\epsilon_\ExcNum(q_+)$ are even functions, the solutions can be chosen to be either even or odd, and the coefficients then fulfill $\gamma_l^{\SubExcNum,\ExcNum} = \pm \gamma_{-l}^{\SubExcNum,\ExcNum}$. It suffices to only consider coefficients with $m \geq 0$ and solve the recurrence equations.

\section{Calculation of perturbation terms}
\label{sec:PerturbativeTerms}

The second order perturbation terms for the potential curves $\tilde \epsilon_\WaveNum(q_+)$ iare found by calculating the matrix elements of the term~(\ref{eq:PertubationTerm}) between pairs of unperturbed eigenfunctions which are plane waves:
\begin{align*}
  I_{l\WaveNum} &\defi - \frac{F(q_+)}{4\pi} \int_{-2\pi}^{+2\pi} \intpar{q_-}  \Exp{\imag(\WaveNum-l)q_-/2} \cos\af{\frac{q_-}{2}} .
\end{align*}
Using the orthogonality of the cosine functions we see that only coefficients with neighboring values of $l$ and $\WaveNum$ are coupled
\begin{align*}
  I_{l\WaveNum} &= - \frac{F(q_+)}{2} [\delta_{\WaveNum-l+1} + \delta_{\WaveNum-l-1}] .
\end{align*}
The resulting perturbative corrections then take the form
\begin{align*}
  \Delta \tilde\epsilon_\WaveNum(q_+)
  &=  \sum_{l \neq \WaveNum}^\infty \frac{\abs{I_{l\WaveNum}}^2}{\tilde\epsilon_\WaveNum - \tilde\epsilon_\SubExcNum}
  = a_\WaveNum F(q_+)^2  
\end{align*}
where the amplitude of the oscillation is
\begin{align*}
  a_\WaveNum &= \sum_{l \neq \WaveNum}^\infty \frac{[\delta_{\WaveNum-l+1} + \delta_{\WaveNum-l-1}]^2}{2 \kappa (\WaveNum^2-l^2) +  16 \pi \gamma (\delta_{\WaveNum,0} - \delta_{l,0})} .
\end{align*}
Here we can distinguish between the three cases $\WaveNum = 0$, $\WaveNum = \pm 1$ and $\abs{\WaveNum} \geq 2$, where we get the results summarized in Eq.~(\ref{eq:CorrectionAmplitudes}).

\bibliography{BornOppenheimerReferencer}

\end{document}